# Valhalla: A Layered Knowledge-State and Service-Governance Framework for Long-Term Scientific Knowledge Work

Yuyang Zheng[1], Nan Li[1], Wenxia Deng[1], Lige Yan[1], Xiang Li[2, *], Si Chen[1, *]

[1]School of Medicine, Shanghai University, Shanghai, China

[2]School of Pharmacy, Second Military Medical University, Shanghai, China

## Abstract

As large language model (LLM) agents are increasingly adopted in scientific research, existing studies have demonstrated that external knowledge bases, knowledge graphs, and long-term memory can improve information retrieval and task continuity. However, most current structured knowledge systems still employ node-centric graph organization, representing files, concepts, experimental results, and knowledge judgments collectively as nodes and relations in a graph. Although this design is suitable for personal knowledge management, in multi-user collaboration and long-term knowledge-transfer settings it tends to make knowledge structures dependent on individual organizational practices, rendering knowledge networks created by different users difficult to share, integrate, and reorganize directly. This paper presents Valhalla, a layered knowledge-state and service-governance framework for long-term scientific knowledge work. The central idea of Valhalla is to transform structured knowledge from a flat graph into a layered encapsulation structure with stable semantic boundaries. The system adopts a five-layer File–Resource–Entity–Relationship–Graph (FREG) knowledge model that separately encapsulates raw-material entry points, source identities, knowledge objects, semantic relations, and task-level knowledge networks, thereby enabling knowledge states produced by different researchers to be exchanged and reorganized according to a unified data structure. Within this model, File and Resource preserve source identity and provenance tracing, Entity objectifies knowledge content, Relationship retains cross-object semantic judgments, and Graph provides task-oriented views for organizing knowledge. On the basis of this layered knowledge state, we further propose a Router–Contract–Workflow service-governance architecture inspired by the microkernel paradigm. This architecture constrains how language models access, modify, and extend knowledge states, ensuring that knowledge sharing maintains structural consistency while also providing auditable operational boundaries. We implement a Valhalla prototype and validate its workflows for knowledge ingestion, cross-member knowledge integration, and scientific writing support through an antibody-design review task. The experiment constructs a structured knowledge state comprising 26 paper resources, 80 knowledge entities, and 92 semantic relations, and demonstrates how knowledge produced by different team members can be organized and reused within a unified layered architecture. The principal contribution of this work is not a new knowledge-extraction algorithm, but a paradigm for organizing structured knowledge in collaborative scientific research: by replacing node-centric flat knowledge graphs with layered encapsulation, Valhalla enables knowledge assets in the LLM era to move beyond individualized organizational schemes and become shared knowledge states with a unified structure that supports transfer and reorganization.

**Github**: https://github.com/fisherrael123-png/Valhalla

# 1 Introduction

As LLM agents have entered scientific and engineering settings, two principal lines of development have emerged.The first extends source acquisition by LLM agents into a process of knowledge organization. This line of research holds that source materials used in scientific and engineering tasks are not themselves directly reusable knowledge, but rather raw carriers of knowledge. Only through entity normalization, evidence binding, relation modeling, and provenance tracing can dispersed, redundant, and context-dependent materials be transformed into stable structured knowledge representations. Such representations allow an LLM agent to move beyond one-off retrieval and short-term context, continually reuse established judgments in subsequent tasks, compare evidence from different sources, maintain dependencies among items of knowledge, and support higher-level reasoning, review, and collaboration [1-3].

The second line focuses on optimizing how LLM agents advance projects and is collectively referred to as AI engineering. It encompasses several subareas, including prompt engineering, retrieval-augmented generation (RAG), context management, skill accumulation and continuous capability extension, and constraint engineering. RAG combines parametric generation with external document retrieval to provide an updatable information interface for knowledge-intensive tasks [4], whereas GraphRAG incorporates cross-document entities and their connections into a graph structure [5]. Research on external memory has, from another perspective, demonstrated the importance of persistent state [6-9]. Context management draws on hierarchical storage and virtual-context management to schedule information between a limited context window and external storage [6]. Skill extension focuses on enabling users to evolve an LLM agent's functionality during use, whereas constraint engineering focuses on advancing projects through independent review mechanisms involving multiple LLM agents [10-15].

Together, these two lines of development establish two essential facts: long-term knowledge work requires both the structuring of content and evidence and the adaptation of LLM agent capabilities to the requirements of knowledge work.

## 2 Main Contributions

In this paper, "long-term scientific knowledge work" refers to research activities in which source materials, judgments, and processes persist across multiple sessions, tasks, or participants. Such work must satisfy the following requirements:

1. Long-term knowledge must be reusable, shareable, and reorganizable for application across diverse scenarios or projects. Reusability is the fundamental purpose of long-term knowledge because it improves an LLM agent's effectiveness; shareability and reorganizability further transform long-term knowledge from mere data into a scientific asset.
2. Core users in scientific teams typically possess substantial domain expertise but may not have backgrounds in systems engineering, software development, or artificial intelligence. Knowledge-management and knowledge-use processes should therefore be readily understandable to users, enabling them to evolve and adjust LLM agents for specific projects.

| Direction | Primary problem addressed | D1 | D2 | D3 | Main remaining gap |
|---|---|---|---|---|---|
| **RAG GraphRAG** | External material retrieval, cross-document semantic organization, and graph-enhanced QA. | ●●○○○ | ●○○○○ | ○○○○○ | Mainly supports information access rather than long-term knowledge governance; lacks low-barrier capability extension for different research tasks. |
| **Long-memory agents and context management** | Context-window extension, external storage, memory scheduling, and cross-session continuity. | ●○○○○ | ●○○○○ | ●○○○○ | Memory units are usually organized around dialogue history or experience fragments, lacking stable, transferable, and auditable scientific knowledge objects. |
| **Skill accumulation self-improvement** | Uses skills, workflows, or accumulated experience to evolve agent capabilities. | ●●○○○ | ●●●●● | ●●○○○ | Provides strong capability extension, but evolved outputs often depend on individual usage traces; lacks structured knowledge-asset sharing and team-level reuse. |
| **Automated scientists and scientific-tool agents** | Automated experiment design, literature analysis, code generation, and research-task planning. | ●●○○○ | ●●○○○ | ●●○○○ | Focuses on automating individual research tasks rather than building a long-term research knowledge environment; lacks cross-project accumulation, transfer, and system governance. |
| **Obsidian + AI** | Human-readable bidirectional notes, material organization, and knowledge networks. | ●●●●○ | ●○○○○ | ○○○○○ | Provides a strong basis for human knowledge organization, but files, sources, concepts, relations, and system configuration are usually not explicitly layered, limiting long-term agent evolution and automated governance. |

| Support level | Low | Low-Mid | Mid | Mid-High | High |
|---|---|---|---|---|---|
| | ●○○○○ | ●●○○○ | ●●●○○ | ●●●●○ | ●●●●● |

*Table 2.1 Relationship between existing LLM-agent research directions and Valhalla's design requirements*

These requirements yield three interrelated design requirements that must nevertheless be validated separately:

D1: Accumulation and transfer of provenance-constrained knowledge states. The system should distinguish file instances, source identities, knowledge objects, semantic relations, and task graphs while preserving paths back to the original materials.

D2: Auditable extension of domain services. The system should represent task entry points, eligibility conditions, and execution steps as explicit components that researchers can read and modify, rather than relying solely on one-off prompts.

D3: Human-in-the-loop long-term governance. Persistent writes, cross-layer modifications to knowledge, and service evolution should be constrained by permissions, state, confirmation, and stage checks, with the location of failures recorded.

These three requirements entail inherent tensions: an excessive emphasis on transfer may erase source context, prioritizing rapid extension may create service coupling, and readable natural-language rules provide weaker guarantees than runtime isolation.

This paper proposes Valhalla, a layered knowledge-state and service-governance framework for long-term scientific knowledge work. Centered on structured knowledge with layered encapsulation, Valhalla constructs as much of the overall system as possible in natural and structured languages, following a microkernel operating-system architecture. Beyond the two principal lines of development introduced above, Valhalla serves two additional purposes. First, layered encapsulation stabilizes knowledge structures, making the sharing of structured knowledge possible. Second, it provides extension mechanisms oriented toward natural language and domain workflows, allowing researchers to adjust knowledge structures, analytical processes, and task capabilities according to the needs of different research projects.

Valhalla is designed to place the trade-offs among D1–D3 within inspectable objects and operational interfaces, rather than to claim that these trade-offs have already been eliminated.

# 3 Overall Architecture of Valhalla

## 3.1 Framework Definition

Valhalla is a layered knowledge-state and service-governance framework for long-term scientific knowledge work. It represents research content as maintainable data objects and requires language models to access or modify those objects through explicit service interfaces. "Content first" means that the system establishes the identities of sources, knowledge, and relations before determining which tasks a model should perform; services may not bypass object boundaries merely for ease of execution.

Analogies to computer systems provide a language for analyzing state, modules, and resource management in LLM-agent architectures. MemGPT likewise uses an operating-system analogy to manage scheduling between context and external memory [6,16]. Valhalla uses this analogy more narrowly: it does not simulate an underlying operating system, but instead adopts concepts such as the data plane, control plane, service entry points, and state checks to organize knowledge operations.

In Valhalla, the long-term knowledge state consists of File, Resource, Entity, Relationship, and Graph. These five object classes assume distinct identity-related responsibilities: File preserves human-facing entry points to source materials; Resource provides a stable source identity; Entity represents a provenance-constrained knowledge object; Relationship records typed semantic judgments; and Graph organizes entities and relations relevant to a task.

## 3.2 Data Plane and Control Plane

The data plane answers the question "What does the system store?", whereas the control plane answers "How does the model operate on these knowledge states?" Neither can substitute for the other. With only a data plane, knowledge objects may still be rewritten arbitrarily by the model; with only a control plane, services lack stable, traceable objects on which to operate.

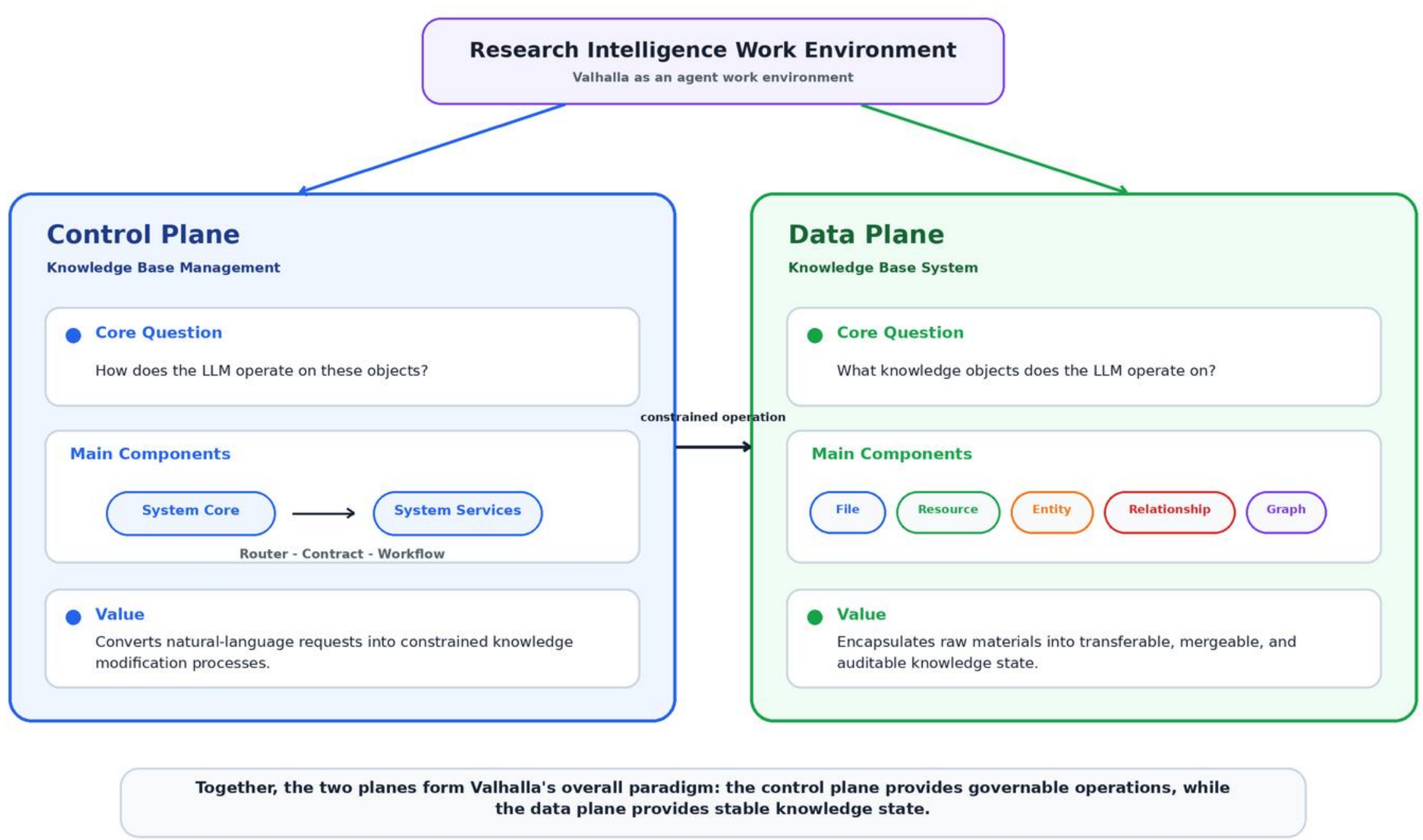

*Figure 3.1 The data plane and control plane jointly constitute Valhalla's overall framework*

Beginning with Files that researchers can manage directly, the data plane establishes stable source identities through Resources, and then separately represents knowledge objects within sources, cross-object judgments, and task views as Entities, Relationships, and Graphs. When evidence must be reviewed, the system traces knowledge objects or relations back to the original materials. When mechanisms or research gaps must be analyzed, it instead prioritizes core semantic relations, avoiding conflation of file locations with knowledge meaning.

The control plane organizes knowledge operations through Router, Contract, and Workflow. Router classifies requests; Contract declares inputs, permissions, risks, system states, confirmation requirements, and output boundaries; and Workflow specifies the execution steps once eligibility conditions have been satisfied. The provenance contracts and skill registry of FundaPod, together with provenance architectures for interactive workflows, provide points of reference for this system-level organization; Valhalla, however, further separates knowledge layers explicitly from operational eligibility [17-19].

Together, the two planes form a design pathway toward D1–D3, rather than evidence that the three objectives have already been achieved. The presence of provenance mappings does not imply that Entities or Relationships are necessarily correct, just as the presence of confirmation rules does not imply that a model cannot circumvent natural-language constraints.

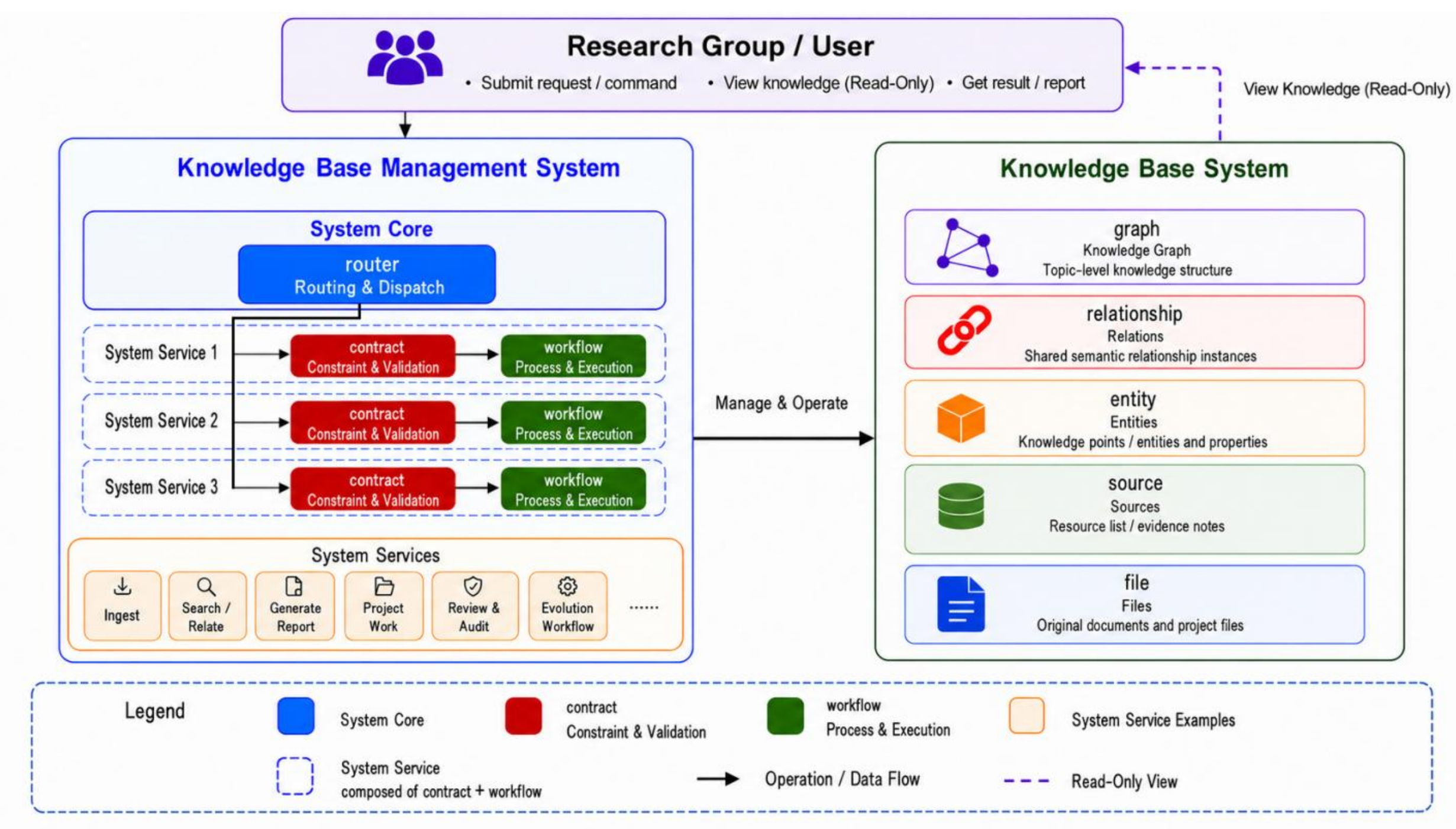

*Figure 3.2 Overall architecture of Valhalla*

# 4 Data Plane: File–Resource–Entity–Relationship–Graph

Valhalla's data-plane design draws on relevant experience from external retrieval, graph structures, evidence graphs, and team knowledge memory. Its focus, however, is not to propose a new extraction algorithm, but to define object identities, maintenance responsibilities, and traceability paths [1,5,20].

The five-layer model can be denoted as $\mathbb{D} \overset{\text{def}}{=} (F, R, E, Rel, \mathcal{G})$, where:

*F*: File, the set of raw files. This file layer is researcher-facing and preserves entry points to the original materials as well as traces of human organization;

*R*: Resource, the set of source materials. This resource layer is system-facing and provides stable source identities and evidence interfaces;

*E*: Entity, the set of knowledge entities. This entity layer objectifies provenance-constrained units of knowledge;

*Rel*: Relationship, the set of relations. This relationship layer records typed semantic judgments across objects;

$\mathcal{G}$: Graph, the set of knowledge graphs. This graph layer organizes relevant entities and relations around specific tasks.

Each layer of the five-layer encapsulation model contains elements with a single type of identity and responsibility, giving the structured knowledge constructed by every user a unified architecture. This makes the model a foundation for sharing structured knowledge.

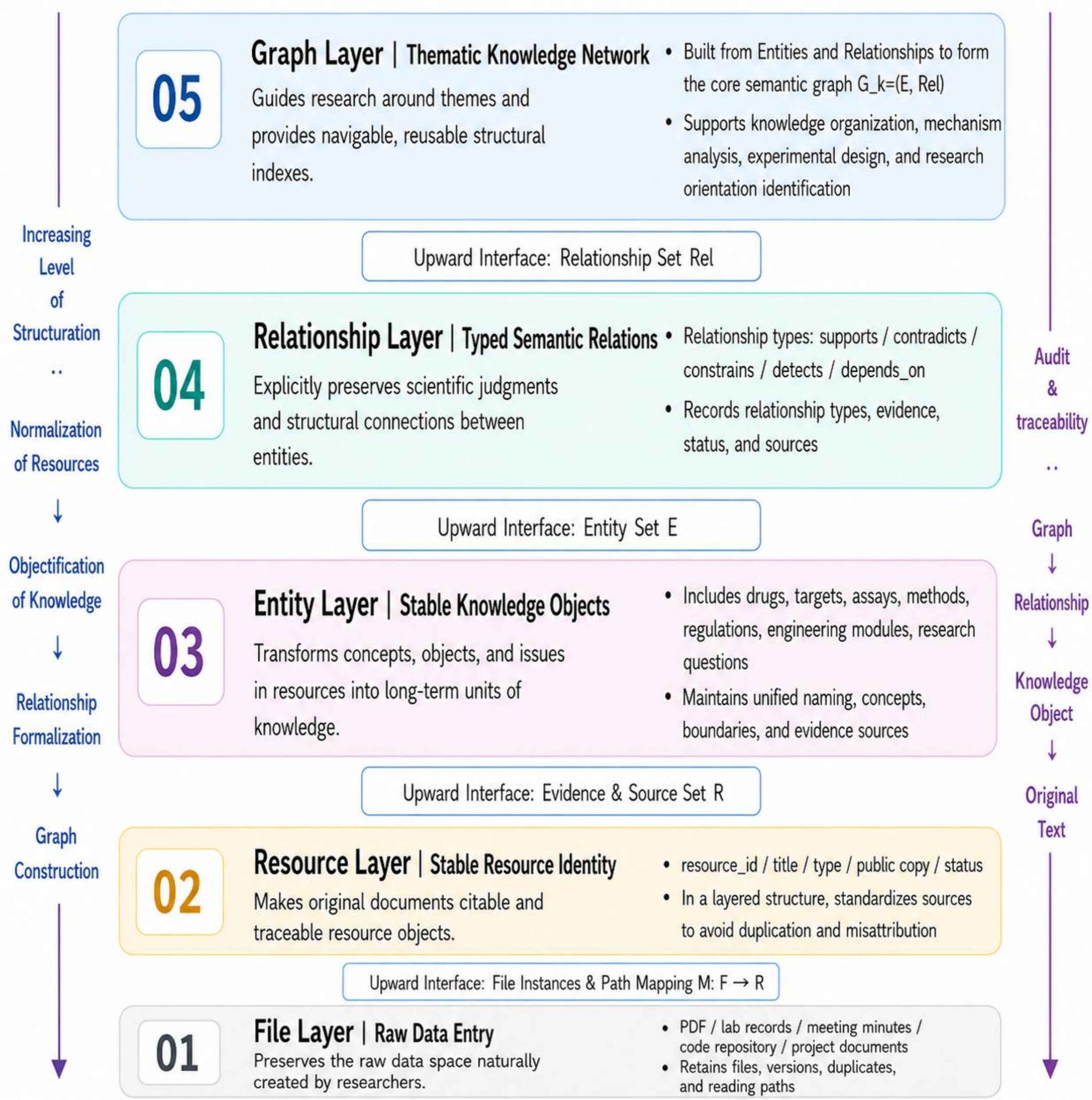

*Figure 4.1 Five-layer object model of the Valhalla knowledge system*

## 4.1 File and Resource: From Human-Facing Entry Points to Stable Source Identities

Scientific materials first enter the workspace in forms such as folders of papers, archived webpages, experimental records, meeting minutes, code repositories, and project documents. Because the location of a file within a directory reflects reading paths, task stages, and human classification, the File layer preserves entry points that researchers can directly understand and maintain.

The File layer does not require physical deduplication through the movement or deletion of files. The same paper may be stored in multiple directories because it was found through different search paths or used in different task contexts; forcibly changing the directory structure would impair the intelligibility of provenance and work processes. Logical normalization should instead occur at the more stable source-identity layer.

The Resource layer abstracts raw files into source objects that the system can identify, reference, and trace. Each Resource has a stable identifier and records the title, type, origin, public copy, lifecycle status, and mappings to File instances. File answers “Where is the source, and why was it placed here?”, whereas Resource answers “How does the system identify the same source?” [2-3].

Accordingly, multiple File instances of the same source may be retained while mapping to a single stable Resource identifier. This mechanism preserves researchers’ file organization while reducing the risk that duplicate files will be mistaken for distinct sources of evidence. Identity matching can nevertheless be erroneous, particularly when versions, supplementary materials, or titles differ, and an interface for manual verification must therefore be retained.

### 4.2 Entity: Provenance-Constrained Knowledge Objects

An Entity represents an independently maintainable knowledge object extracted from a source, rather than the source file itself. Real-world literature simultaneously contains background information, methods, experimental conditions, author conjectures, and noise. The purpose of objectification is to delineate the boundaries of reusable knowledge units while preserving their provenance.

A Resource may contain multiple relatively independent problems, mechanisms, or methods. If these elements have distinct evidence locations, scopes of applicability, or conclusions, they should form separate Entities rather than being compressed into a single summary. The granularity of decomposition must support subsequent tracing and relation construction; it cannot be determined solely by titles or keywords.

Similar knowledge from different sources is not merged at this layer. Similar statements may depend on different experimental conditions and argumentative contexts, and direct merging would erase differences in provenance. Valhalla retains source-specific Entities separately and uses Relationships to express equivalence, overlap, support, attenuation, or conflict [1-3].

The Entity layer therefore performs the transformation from source materials into maintainable knowledge objects. It retains evidentiary provenance through Resource but does not independently perform integration across sources. Object boundaries and content correctness still require review, and an active status in the system cannot substitute for domain-level judgments of factual validity.

### 4.3 Relationship: Typed Semantic Judgments

The reuse of scientific knowledge depends not only on individual objects but also on inspectable connections among them. A Relationship contains, at minimum, a source node, a target node, a relation type, a textual description, a status, and evidentiary provenance. Relation types may be defined according to the task and may include supports, contradicts, causes, constrains, detects, implements, and depends_on [1,3].

Explicit relations allow the system to determine whether a conclusion has a chain of support, whether a proposed approach contains gaps in its evidence, and which dependent modules may be affected by an engineering change. Their value lies in preserving the structure of judgments and providing entry points for tracing, rather than in substituting lines in a graph for causal proof.

## 4.4 Graph: Task-Oriented Knowledge Networks

A Graph consists of selected Entities and Relationships and may correspond to a research topic, project, review, experimental protocol, or engineering module. It differs from file-link graphs in systems such as Obsidian: nodes in a file-link graph span both the entity and file layers, and its links likewise cross layer boundaries [21]. By contrast, Valhalla's core semantic graph presents objectified knowledge and typed judgments.

The role of a Graph is to provide a structured index for the current task, rather than to visualize every object in the knowledge base. Researchers can first locate key Entities, Relationships, and evidence gaps, and then trace them back to the original sources as needed. Insufficient graph coverage or numerous isolated nodes should be treated as gaps in knowledge organization, rather than as a complete theoretical structure.

## 4.5 Core Semantic Graph and Evidence Mappings

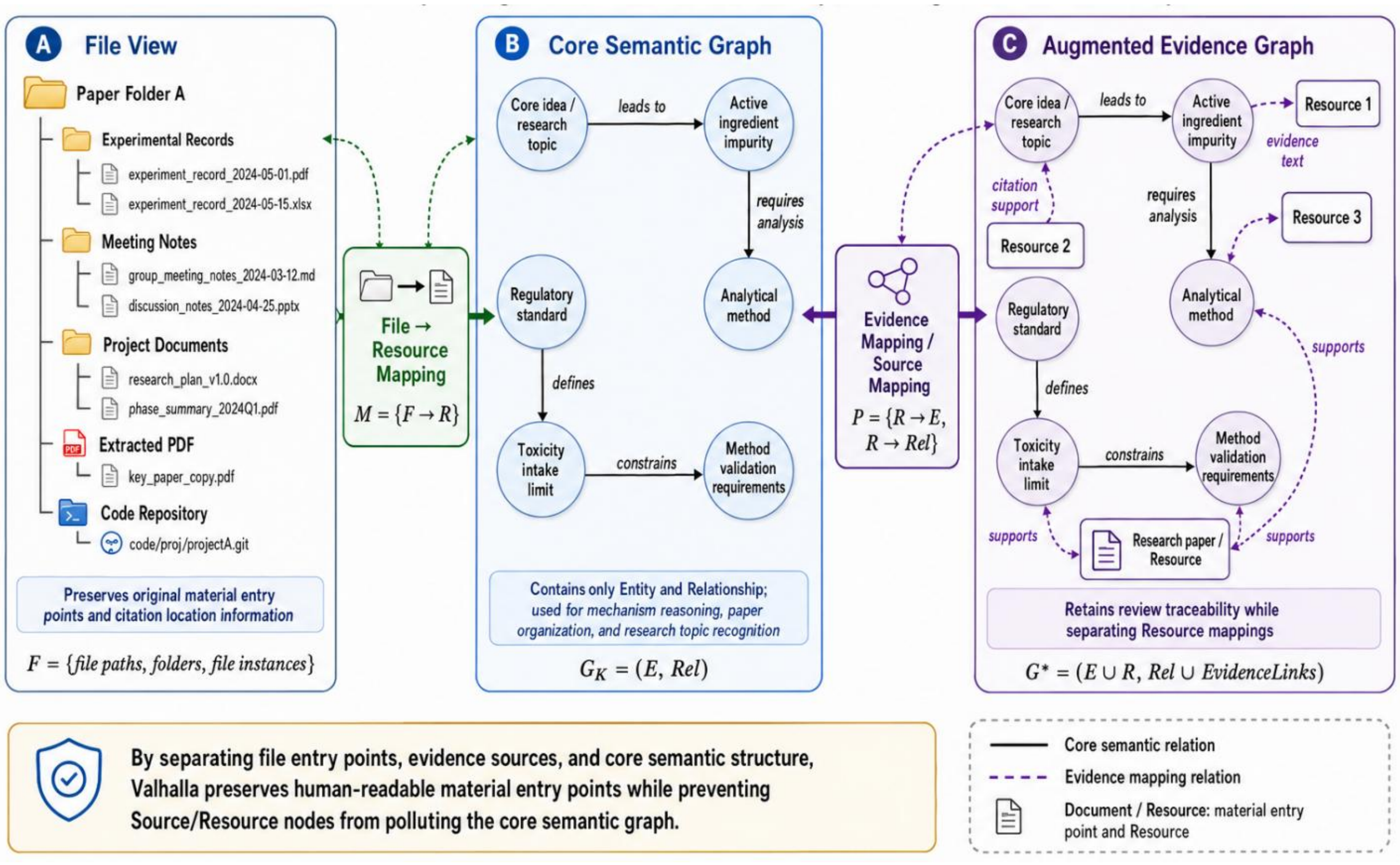

*Figure 4.2 Structural relationships among the file view, core semantic graph, and extended evidence graph*

To avoid conflating file locations, source provenance, and conceptual relations, Valhalla distinguishes among the file view, core semantic graph, and extended evidence graph. The file view retains the directory structure of Files; the core semantic graph contains only Entities and Relationships; and the extended evidence graph adds Resource mappings when verification is required, thereby connecting knowledge judgments to the original sources.

The system can switch perspectives according to the task: users inspect the core semantic graph to understand knowledge structure, expand Resource mappings to review provenance, and locate original files through File–Resource mappings when returning to source materials. Such switching reduces the conflation of views but cannot compensate for erroneous source registration, entity extraction, or relation judgments.

# 5 Formal Architecture of the Data Plane $\mathbb{D}$

## 5.1 Five-Layer Knowledge-State Architecture

$$\mathbb{D} \stackrel{\text{def}}{=} (F, R, E, Rel, \mathcal{G})$$

A concrete implementation of $\mathbb{D}$ in the data plane is referred to as Root $\rho$; thus:

$$\mathbb{D}_\rho \stackrel{\text{def}}{=} \left(F_\rho, R_\rho, E_\rho, Rel_\rho, \mathcal{G}_\rho\right)$$

where:

$F_\rho$: the set of Files in Root $\rho$.

$R_\rho$: the set of Resources in Root $\rho$.

$E_\rho$: the set of Entities in Root $\rho$.

$Rel_\rho$: the set of Relationships in Root $\rho$.

$\mathcal{G}_\rho$: the set of Graphs in Root $\rho$.

## 5.2 Inter-Layer Knowledge-State Transformations in the Data Plane, $\mathcal{M}^t$

The five layers of the Valhalla data plane are connected through a set of mappings with distinct responsibilities that perform inter-layer knowledge-state transformations (Inter-layer knowledge-state transformations of the data plane).

The five encapsulated structured-data layers are linked by four transformation mappings that serve as interfaces between adjacent layers.

**1. File-to-Resource Transformation Mapping $\mathcal{M}_{FR}$**

This mapping maps an original File to a standardized logical Resource.

$$\mathcal{M}_{FR}^t: \{f \mapsto r\}$$

where:

$$f \in F, r \in R$$

Multiple Files may be mapped to a single Resource:

$$f_1 \neq f_2, \quad \mathcal{M}_{FR}^t(f_1) = \mathcal{M}_{FR}^t(f_2) = r$$

For example, File instances with different paths and filenames, or even instances saved by different members, may be identified by the system as the same source material.

This can be represented as a set mapping (function):

$$\mathcal{M}_{FR}^{t}: \{F \mapsto R\}$$

**2. Resource-to-Entity Transformation Mapping $\mathcal{M}_{RE}$**

Entities are extracted from a Resource.

$$\mathcal{M}_{RE}^{t}: \{r \mapsto e\}$$

A Resource may contain multiple Entities:

$$\mathcal{M}_{RE}^{t}: \{r \mapsto e_1, e_2, e_3, \cdots\}$$

This can be represented as a set mapping (function):

$$\mathcal{M}_{RE}^{t}: \{R \mapsto \mathcal{P}(E)\}$$

$\mathcal{P}(E)$: the power set of $E$.

**3. Entity-to-Relationship Transformation Mapping $\mathcal{M}_{ERel}$**

Relationships among Entities are summarized and organized.

$$\mathcal{M}_{ERel}^{t}: \{e_1, e_2 \mapsto \ell\}$$

Two Entities may have Relationships of different types:

$$\mathcal{M}_{ERel}^{t}: \{e_1, e_2 \mapsto \ell_1, \ell_2, \cdots\}$$

This can be represented as a set mapping (function):

$$\mathcal{M}_{ERel}^{t}: \{E \times E \mapsto \mathcal{P}(Rel)\}$$

$\times$: Cartesian product.

$\mathcal{P}(Rel)$: the power set of $Rel$.

**4. Relationship-to-Graph Transformation Mapping $\mathcal{M}_{RelG}$**

Distinct knowledge graphs are constructed according to topic.

$$\mathcal{M}_{RelG}^{t}: \{\ell_1, \ell_2, \cdots \mapsto g\}$$

This can be represented as a set mapping (function):

$$\mathcal{M}_{RelG}^{t}: \{\mathcal{P}(Rel) \mapsto \mathcal{P}(\mathcal{G})\}$$

Because $\mathcal{G}$ also records information about Entities:

$$\mathcal{M}_{RelG}^{t}: \{\mathcal{P}(E) \times \mathcal{P}(Rel) \mapsto \mathcal{P}(\mathcal{G})\}$$

$\times$: Cartesian product.

$\mathcal{P}(Rel)$: the power set of $Rel$.

$\mathcal{P}(E)$: the power set of $E$.

**5. Composite Transformation Mappings**

The four transformation mappings can be summarized as:

$$\mathcal{M}^{t}: \{\mathcal{P}(F) \mapsto \mathcal{P}(R); \mathcal{P}(R) \mapsto \mathcal{P}(E); \mathcal{P}(E) \mapsto \mathcal{P}(Rel); \mathcal{P}(Rel) \mapsto \mathcal{P}(\mathcal{G})\}$$

Adjacent mappings can be composed. For example:

$$\mathcal{P}(Rel) = \mathcal{M}^t_{ERel}(\mathcal{M}^t_{RE}(\mathcal{M}^t_{FR}(F))) = \mathcal{M}^t_{ERel} \circ \mathcal{M}^t_{RE} \circ \mathcal{M}^t_{FR}(F)$$

For readability, this may also be expressed as:

$$\mathcal{P}(Rel) = \mathcal{M}^t_{FR}\mathcal{M}^t_{RE}\mathcal{M}^t_{ERel}(F)$$

Similarly:

$$\mathcal{P}(\mathcal{G}) = \mathcal{M}^t_{RelG}(\mathcal{M}^t_{ERel}(\mathcal{M}^t_{RE}(\mathcal{M}^t_{FR}(F)))) = \mathcal{M}^t_{FR}\mathcal{M}^t_{RE}\mathcal{M}^t_{ERel}\mathcal{M}^t_{RelG}(F)$$

## 5.3 Provenance Mapping, $\mathcal{M}^p$

Transformation mappings construct the five encapsulated layers of structured knowledge from the bottom upward, layer by layer. Provenance mapping proceeds from the top downward to map each object to its sources [3].

**1. Graph Provenance Mapping $\mathcal{M}^p_G$**

Although a Graph is constructed from Relationships, it records both the Relationships and the Entities used to construct it.

$$\mathcal{M}^p_G: \{g \mapsto \ell_1, \ell_2, \cdots, g \mapsto e_1, e_2, e_3, \cdots\}$$

Represented as a set mapping (function):

$$\mathcal{M}^p_G: \{\mathcal{P}(\mathcal{G}) \mapsto \{\mathcal{P}(Rel), \mathcal{P}(E)\}\}$$

**2. Relationship Provenance Mapping $\mathcal{M}^p_{Rel}$**

A Relationship records the Entities from which it was constructed.

$$\mathcal{M}^p_{Rel}: \{\ell \mapsto e_1 \times e_2\}$$

Represented as a set mapping (function):

$$\mathcal{M}^p_{Rel}: \{\mathcal{P}(Rel) \mapsto \mathcal{P}(E)\}$$

**3. Entity Provenance Mapping $\mathcal{M}^p_E$**

An Entity records the Resource from which it was constructed.

$$\mathcal{M}^p_E: \{e \mapsto r\}$$

Represented as a set mapping (function):

$$\mathcal{M}^p_E: \{\mathcal{P}(E) \mapsto \mathcal{P}(R)\}$$

**4. Resource Provenance Mapping $\mathcal{M}^p_R$**

A Resource records the Files from which it was constructed.

$$\mathcal{M}^p_R: \{r \mapsto f_p, f_1, f_2, \cdots\}$$

Here, $f_p$ is the common copy of $f_1, f_2, \cdots$. This common copy is established so that $\mathcal{M}^p_R$ can obtain a canonical record $f_p$ unaffected by manual operations, thereby preventing changes to Files from rendering provenance unreachable.

Represented as a set mapping (function):

$$\mathcal{M}_R^p: \{\mathcal{P}(R) \mapsto \{\mathcal{P}(F_p), \mathcal{P}(F)\}\}$$

**5. Composite Provenance Mappings**

The four provenance mappings can be summarized as:

$$\mathcal{M}^p: \{\mathcal{P}(\mathcal{G}) \mapsto \{\mathcal{P}(Rel), \mathcal{P}(E)\}; \mathcal{P}(Rel) \mapsto \mathcal{P}(E); \mathcal{P}(E) \mapsto \mathcal{P}(R); \mathcal{P}(R) \mapsto \{\mathcal{P}(F_p), \mathcal{P}(F)\}\}$$

Adjacent mappings can be composed. For example:

$$\mathcal{P}(R) = \mathcal{M}_E^p(\mathcal{M}_G^p(\mathcal{G})) = \mathcal{M}_E^p \circ \mathcal{M}_G^p(\mathcal{G})$$

For readability, this can be written as:

$$\mathcal{P}(R) = \mathcal{M}_G^p \mathcal{M}_E^p(\mathcal{G})$$

Consequently, when knowledge Graph G is used, its provenance can, where necessary, be traced back to the original Files.

## 5.4 Core Semantic Graph and Extended Evidence Graph

The core semantic graph for topic T is:

$$\mathbb{G}_T \stackrel{\text{def}}{=} \{Rel_T, E_T\}$$

where:

$$E_T \subseteq E_\omega, Rel_T \subseteq Rel_\omega$$

The extended evidence graph for topic T is:

$$\mathbb{G}_T^e \stackrel{\text{def}}{=} \{Rel_T, E_T, R_T, F_T\}$$

where:

$$R_T = \mathcal{M}_E^p(E), F_T = \mathcal{M}_E^p \mathcal{M}_R^p(E)$$

That is:

$$\mathbb{G}_T^e = \mathbb{G}_T \cup \mathcal{M}_G^p \mathcal{M}_E^p(\mathbb{G}_T) \cup \mathcal{M}_G^p \mathcal{M}_E^p \mathcal{M}_R^p(\mathbb{G}_T)$$

# 6 Formal Architecture of the Implemented Data Plane $\mathbb{D}^*$

$\mathbb{D}$ provides the architectural foundation for sharing structured knowledge, but the reorganization of structured knowledge must also be implemented.

## 6.1 Implemented Knowledge-State Architecture $\mathbb{D}^*$

In practical implementation, the five-layer knowledge-state architecture supports the reuse and reorganization of source materials at the Resource layer through the construction of multiple knowledge bases.

$$\mathbb{D}^* \stackrel{\text{def}}{=} (F, R, \Omega)$$

The knowledge-base set $\Omega$ comprises one or more knowledge bases $\omega$:

$$\Omega = \{\omega_1, \omega_2, \cdots\}$$

Each knowledge base $\omega$ has an independent knowledge space composed of $\{E, Rel, \mathcal{G}\}$. In addition, each knowledge base has a virtual Resource layer V to support constrained provenance tracing.

$$\omega \stackrel{\text{def}}{=} \{V_\omega, E_\omega, Rel_\omega, \mathcal{G}_\omega\}$$

Therefore:

$$\mathbb{D}^* \stackrel{\text{def}}{=} \left(F, R, \{\{V, E, Rel, \mathcal{G}\}_{\omega_1}, \{V, E, Rel, \mathcal{G}\}_{\omega_2}, \cdots\}\right)$$

## 6.2 Virtual Resource Layer $V_\omega$

The virtual Resource layer $V_\omega$ consists of four components, all of whose entries are Resources rather than Files.

$L_\omega$: the local Resource table, local_resources. This table contains materials for general use. When knowledge base $\omega$ is used to advance a project, materials registered here are included within the scope of consideration but are not necessarily used.

$Q_\omega$: the required Resource table, required_resources. When knowledge base $\omega$ is used to advance a project, materials registered here must be used.

$X_\omega$: the excluded Resource table, excluded_resources. When knowledge base $\omega$ is used to advance a project, materials registered here must not be used. This table has a higher priority than $Q_\omega$; that is, when $X_\omega$ conflicts with $Q_\omega$, the entries in $X_\omega$ are retained.

$B$: the blacklist of a given Root. It contains materials globally disabled within that Root; for example, if a source is found to involve academic fraud, it is excluded from the knowledge system through the blacklist mechanism.

Thus:

$$V_\omega = (L_\omega \cup Q_\omega) \setminus \left(X_\omega \cup B_\rho\right)$$

The actual Resource set in use, $A_\omega$, then satisfies:

$$Q_\omega \setminus \left(X_\omega \cup B_\rho\right) \subseteq A_\omega \subseteq V_\omega$$

## 6.3 Resource-to-$V$ Transformation Mapping $\mathcal{M}_{RV}^t$

$$\mathcal{M}_{RV}^t = \{\mathcal{M}_{RL}^t, \mathcal{M}_{RQ}^t, \mathcal{M}_{RX}^t\}$$

Resources R in $\rho$ are mapped to entries in the local Resource table $L_\omega$:

$$\mathcal{M}_{RL}^t: \{R \mapsto L_\omega\}$$

Resources R in $\rho$ are mapped to entries in the required Resource table $Q_\omega$:

$$\mathcal{M}_{RQ}^t: \{R \mapsto Q_\omega\}$$

Resources R in $\rho$ are mapped to entries in the excluded Resource table $X_\omega$:

$$\mathcal{M}_{RX}^t: \{R \mapsto X_\omega\}$$

## 6.4 $V$-to-Entity Transformation Mapping $\mathcal{M}_{VE}^{t}$

$$\mathcal{M}_{VE}^{t}: \{A_\omega \mapsto E_\omega\}$$

It therefore follows that:

$$\mathcal{G}_\omega = \mathcal{M}_{FR}^{t}\mathcal{M}_{RV}^{t}\mathcal{M}_{VE}^{t}\mathcal{M}_{ERel}^{t}\mathcal{M}_{RelG}^{t}(F)$$

## 6.5 Shared Data Plane

Root fusion is used as an example to illustrate the role of the virtual Resource layer $V_\omega$.

Let the two distinct Roots be:

$$\rho_1, \rho_2$$

They are fused into a new Root:

$$\rho_3$$

**1. Resource-Layer Integration**

The Resource layer of $\rho_3$ is:

$$R_3 = Normalize(R_1 \cup R_2)$$

where:

Normalize denotes:

- the reassignment of unified identities in the Resource layer;

- the continued separate retention of different versions of the same publication;

- the avoidance of erroneous merging between similar but distinct materials.

**2. Blocking Strategy**

For a Resource r subject to a blocking conflict between $B_1, B_2$—that is, r is blocked by one but not the other—a user decision is defined as:

$$\mu(r) \in \{global, local, allow\}$$

where:

$global$: the Resource is globally blocked in the new Root by adding it to $B_3$;

*local*: the Resource is blocked only locally within the relevant knowledge bases. Specifically, it is not added to $B_3$; instead, under the Root whose $B_\rho$ originally contained an entry for r, the knowledge-base set $\Omega$ has r added to the applicable $X_\omega$ entries, thereby implementing local blocking. That is:

$$X_\omega^{*} = X_\omega \cup \{r | \mu(r) = local\}$$

*allow*: the Resource is no longer blocked.

**3. Fused Knowledge Base $\Omega$**

The knowledge bases $\Omega$ under the two Roots are fused according to the blocking strategy.

# 7 Control Plane: Constrained Knowledge Operations

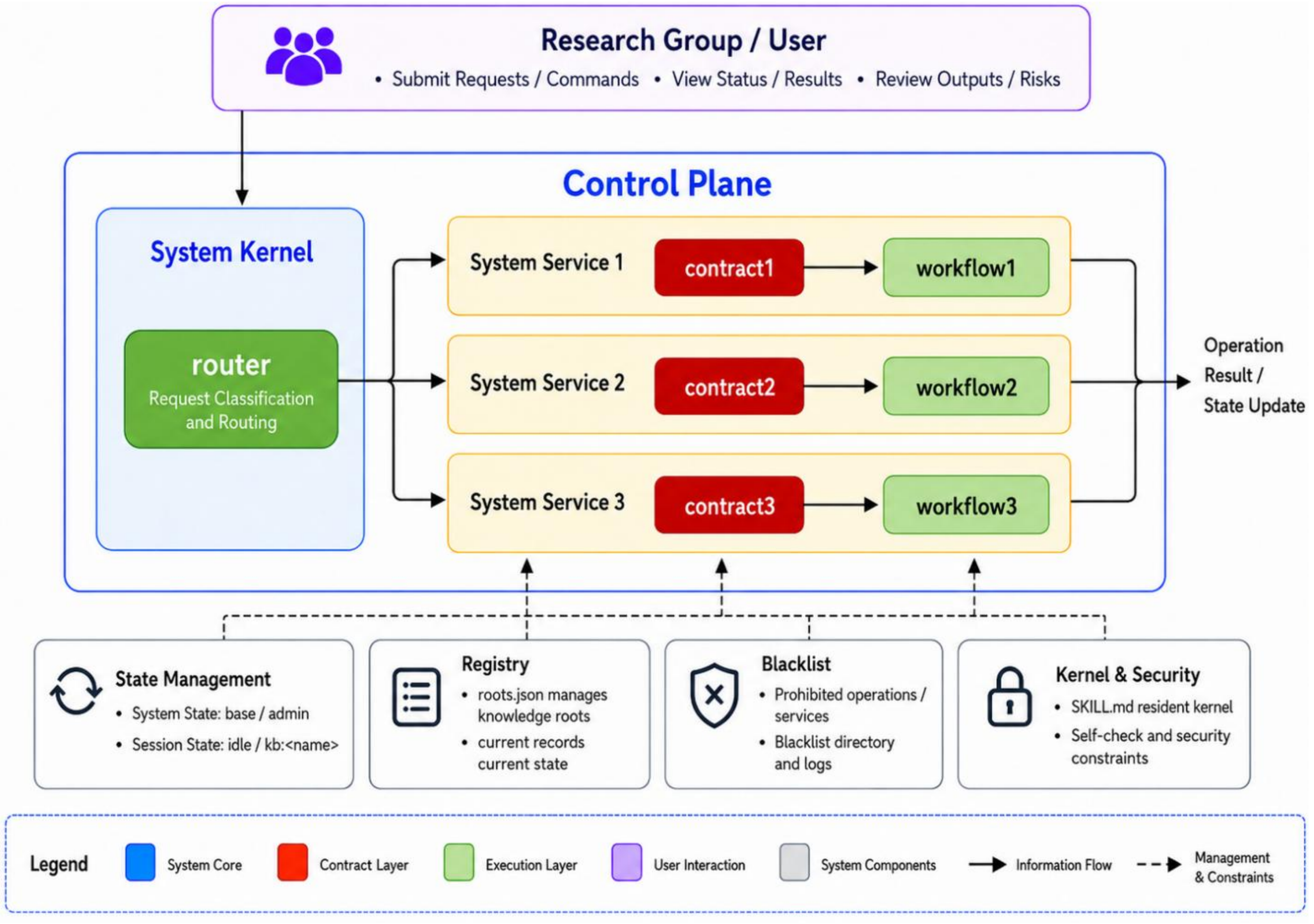

*Figure 7.1 The control plane for knowledge operations in Valhalla*

Valhalla's control plane transforms natural-language requests into system services subject to explicit eligibility conditions. Its purpose is to make request classification, access scope, risk, confirmation, and output validation explicit, rather than to simulate a formal authorization system through natural-language rules [22-23].

## 7.1 Microkernel-Based Service Structure

Valhalla treats the language model as a processor, the current conversational context as an in-memory workspace, and long-term knowledge and service definitions as programs and data in external storage. Inspired by hierarchical memory and computer systems, this analogy explains the use of minimal resident rules and on-demand service loading; it does not imply that the system already provides process isolation or memory protection [6,16,23].

Valhalla is implemented as an explicitly loadable skill, allowing the same governance interface to be used across different LLM agent environments. The implementation vehicle itself is not presented as a performance contribution; its purpose is to keep system states, service entry points, and safety boundaries readable.

After initialization, top-level rules are responsible for self-checks, state identification, service routing, and baseline safety constraints. They neither store the entirety of scientific knowledge nor load all workflows at once. Their role approximates the minimal resident interface of a microkernel: maintaining only the rules required for service selection and boundary checking.

Specific capabilities are loaded on demand. Following operating-system terminology, this capability-execution process is referred to as a system service.

A system service can be represented as $\mathcal{s} \stackrel{\text{def}}{=} \{\mathcal{R}, \mathcal{C}, \mathcal{W}\}$

where:

$\mathcal{R}$: Router, for request routing;

$\mathcal{C}$: Contract, the operational contract;

$\mathcal{W}$: Workflow, the workflow.

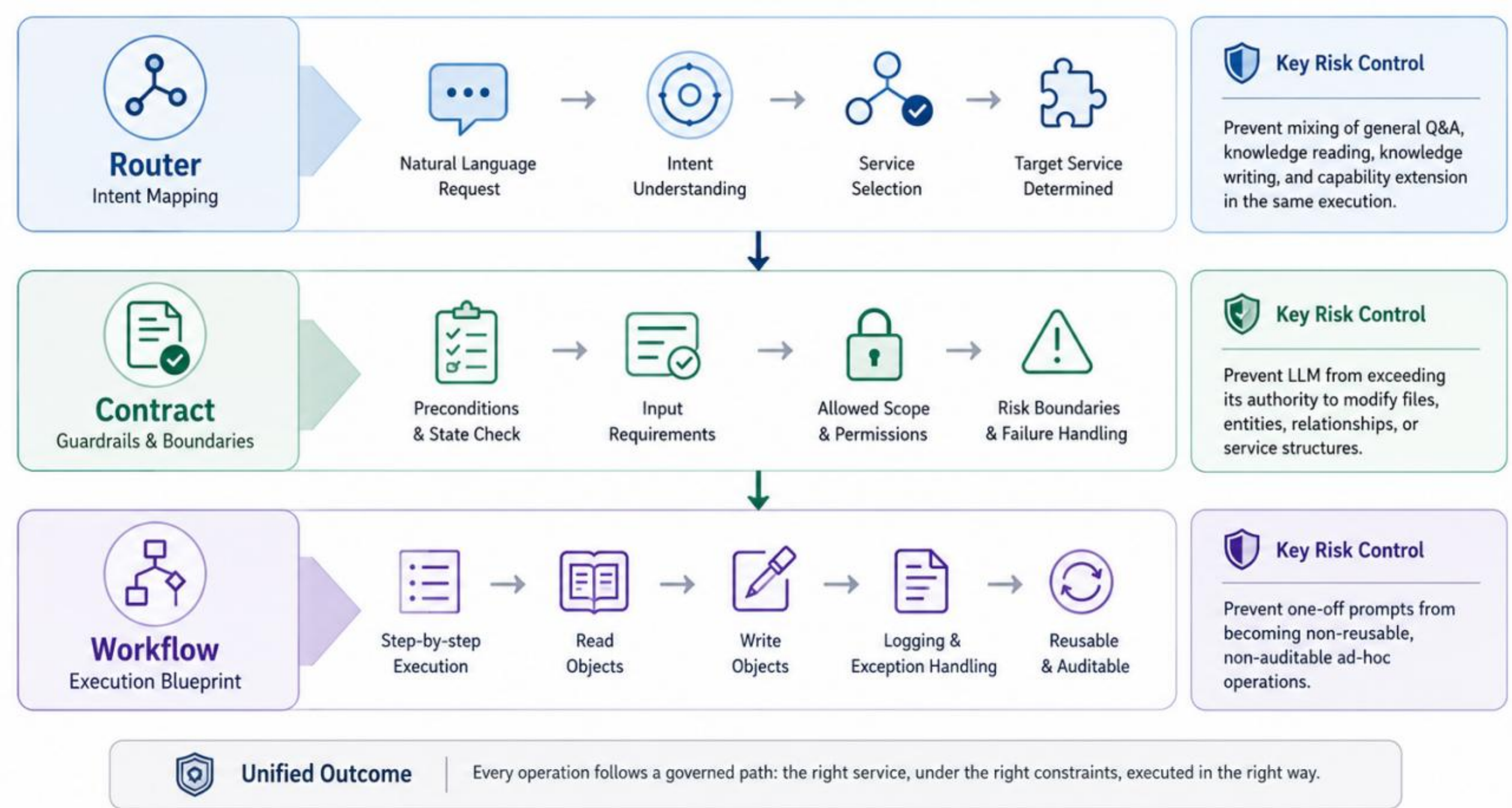

*Figure 7.2 The system service of Valhalla*

The "minimal residency + on-demand loading" design reduces the extent to which irrelevant rules occupy the context and facilitates the localization of service changes. However, the current implementation still relies on context management by the underlying model and cannot guarantee that critical constraints will always be preserved intact after long-session context compression [24].

## 7.2 Router: Routing Natural-Language Requests to Services

Researchers typically use natural language to request resource ingestion, entity maintenance, relationship construction, graph organization, paper writing, or system extension. Although intuitive to humans, these expressions involve different objects, permissions, and write scopes. The Router is responsible for mapping each request to an explicit operation category.

The Router first distinguishes among general question answering, knowledge-base reading, knowledge ingestion, structural maintenance, project work, and service extension, and then selects the corresponding Contract. If a request matches multiple operations or its target is unclear, the system should expose the ambiguity and request clarification, rather than autonomously combining multiple write scopes.

The Router neither modifies knowledge nor determines whether a request is ultimately authorized. It serves as the service-dispatch entry point, reducing the risk that the model will directly operate on File, Resource, Entity, Relationship, or Graph objects based solely on surface semantics.

## 7.3 Contract: Operational Eligibility and Service Contract

The Contract first determines whether an operation may be executed under the current conditions. It then authorizes the operation by specifying the required inputs, read and write permissions, risk level, permitted states, prerequisites, confirmation requirements, access scope, required outputs, and failure-handling procedures. Because different services operate on different objects, they cannot share a vague, generic authorization. Finally, the Contract provides the entry point to the Workflow [19,22-23].

If a Workflow delegates execution to another service, the target Contract should perform its checks again rather than inherit confirmation from the preceding service.

The Contract converts implicit requirements that depend on the model's voluntary compliance into readable boundaries, allowing reviewers to determine what the system intends to read or write, why confirmation is required, and what should be returned upon failure. Nevertheless, it remains a normative constraint rather than an unbypassable security guarantee.

## 7.4 Workflow: Readable and Modifiable Execution Procedures

Only after the Router has classified the request and the Contract has passed its eligibility checks does the system load the Workflow. The Workflow specifies the order of reads, registry checks, state updates, logging, output validation, and conflict handling, so that tasks of the same type need not be replanned from an ad hoc prompt each time [18-19].

A literature-ingestion workflow may specify that the system first establish the identity of a resource, extract candidate knowledge objects, check existing registrations, write source mappings, and run validation. A relationship-construction workflow may specify that the system first verify object identities, select relationship types, bind original evidence, and update the task graph. The value of these workflows lies in their reusability and auditability, not in any guarantee that the model will never omit a step.

Natural language and lightweight structured files lower the barrier for interdisciplinary teams to modify workflows, but they also introduce ambiguity and refactoring risks. Consequently, critical isolation, transactional commits, and permission enforcement should still be implemented by external orchestrators or code-based mechanisms rather than relying exclusively on Workflow text [22-23].

# 8 Semi-Formal Architecture of the Control Plane $\mathbb{C}$

## 8.1 Microkernel Architecture

$$\mathbb{C} \stackrel{\text{def}}{=} \{\mathcal{K}, \mathcal{S}\} = \{\mathcal{K}, \{s_1, s_2, s_3, \cdots\}\}$$

where:

$\mathcal{K}$: the system kernel of the control plane $\mathbb{C}$;

$\mathcal{S}$: the set of system services in the control plane $\mathbb{C}$, $\mathcal{S} = \{s_1, s_2, s_3, \cdots\}$;

$s$: a system service.

## 8.2 System Kernel $\mathcal{K}$

$$\mathcal{K} = \{bootstrap,\ \ security,\ \ rule\}$$

where:

$bootstrap$: startup initialization, implemented in Python and executed as a mandatory step, including stable foundational operations such as checking system integrity, resetting system states, and loading the Router;

$security$: safety constraints;

$rule$: global rules.

## 8.3 System State $\mathbb{V}$ and Session State $\mathbb{S}$

Valhalla employs two state models: the system state $\mathbb{V}$ and the session state $\mathbb{S}$.

### 1. System State $\mathbb{V}$

This state describes the overall state of Valhalla and is used to manage operational permissions.

$$\mathbb{V} = \begin{cases} base & \text{Base state, in which basic operations may be performed} \\ admin & \text{Administrative state, in which administrative operations may be performed} \end{cases}$$

### 2. Session State $\mathbb{S}$

This state describes the current Valhalla session and identifies the target of system services.

$$\mathbb{S} = \begin{cases} idle & \text{Idle state; no service target} \\ kb\text{: } < kb\ name > & \text{Active state; service target: } < kb\ name > \end{cases}$$

## 8.3 System Service $s$

$$s \stackrel{\text{def}}{=} \{\mathcal{R}, \mathcal{C}, \mathcal{W}\}$$

where:

$\mathcal{R}$: Router, for request routing;

$\mathcal{C}$: Contract, the operational contract;

$\mathcal{W}$: Workflow, the workflow.

The transformation mapping $\mathcal{M}^t$ introduced in Chapter 5 is instantiated here as multiple system services $\mathcal{s}$.

## 8.4 Request-Routing Component: Router

The Router maps a user's natural-language request to the entry point of the corresponding type of operational contract.

$$\mathcal{R} \stackrel{\text{def}}{=} \{pattern, key, \mathcal{C} - entrance\}$$

where:

$pattern$: request type;

$key$: keywords that trigger the corresponding Router entry;

$\mathcal{C} - entrance$: the operational-contract entry point for the request.

## 8.5 Operational Contract: Contract

The Contract uses structured language to constrain workflow execution.

$$\mathcal{C} \stackrel{\text{def}}{=} \{qualification_test, target, permission, \mathcal{W} - entrance\}$$

where:

$qualification_test$: determines whether the system service may be executed under the current system state $\mathbb{V}$ and session state $\mathbb{S}$;

$target$: determines the target of the current system service according to the current session state $\mathbb{S}$;

$permission$: specifies the permissions available to the current system service, such as write permission and read-access scope;

$\mathcal{W} - entrance$: the entry point to the Workflow.

## 8.6 Workflow

A workflow constructed in natural language.

## 8.7 System Service $\mathcal{s}_1$ Calling System Service $\mathcal{s}_2$

To preserve the system's microkernel architecture, system services should be decoupled from one another. For convenience, however, a very small number of system services may call other system services.

Under the global rules in $\mathcal{K}$, when system service $\mathcal{s}_1$ calls system service $\mathcal{s}_2$, it must execute $\mathcal{s}_2$ in full and may not directly execute $\mathcal{W}_2$. That is,

$$\mathcal{s}_1 = \{\mathcal{R}_1, \mathcal{C}_1, \mathcal{W}_1^{step1}, \mathcal{s}_2, \mathcal{W}_1^{step2}\}$$

Because $\mathcal{s}_2 = \{\mathcal{R}_2, \mathcal{C}_2, \mathcal{W}_2\}$, this is equivalent to:

$$\mathcal{s}_1 = \{\mathcal{R}_1, \mathcal{C}_1, \mathcal{W}_1^{step1}, \{\mathcal{R}_2, \mathcal{C}_2, \mathcal{W}_2\}, \mathcal{W}_1^{step2}\}$$

It is not equivalent to bypassing the second service's Router and Contract:

$$\mathcal{s}_1 \neq \{\mathcal{R}_1, \mathcal{C}_1, \mathcal{W}_1^{step1}, \mathcal{W}_2, \mathcal{W}_1^{step2}\}$$

### 8.8 Idempotent Operations

Valhalla supports idempotent operations: repeatedly executing a system service $\mathcal{s}$ does not further change any knowledge state $\mathbb{S}$.

$$\mathcal{s}(\mathcal{s}(\mathbb{S})) = \mathcal{s}(\mathbb{S})$$

## 9 Human-in-the-Loop Governed Service Evolution

### 9.1 Definition

In this paper, "governed service evolution" refers to extending or modifying system services, through either manual evolution or self-evolution, after a user identifies a need for functional change. Self-evolution means that the LLM evolves services in accordance with predefined rules.

The basic rules of service evolution are as follows:

1. $\mathcal{s}$ follows the fixed architecture $\{\mathcal{R}, \mathcal{C}, \mathcal{W}\}$;
2. $\mathcal{s}$ has an explicit $\mathcal{R}$ component $\{pattern, key, \mathcal{C} - entrance\}$;
3. $\mathcal{s}$ has an explicit $\mathcal{C}$ component $\{qualification_test, target, permission, \mathcal{W} - entrance\}$;
4. $\mathcal{s}$ has an explicit target. If it targets structured knowledge maintained by Valhalla, the target must be explicitly identified as one or more layers within the layered encapsulation $\mathbb{D}^*$;
5. The workflow $\mathcal{W}$ of $\mathcal{s}$ is constructed in natural language to facilitate review of the system's operating mechanisms.

### 9.2 Evolution Process

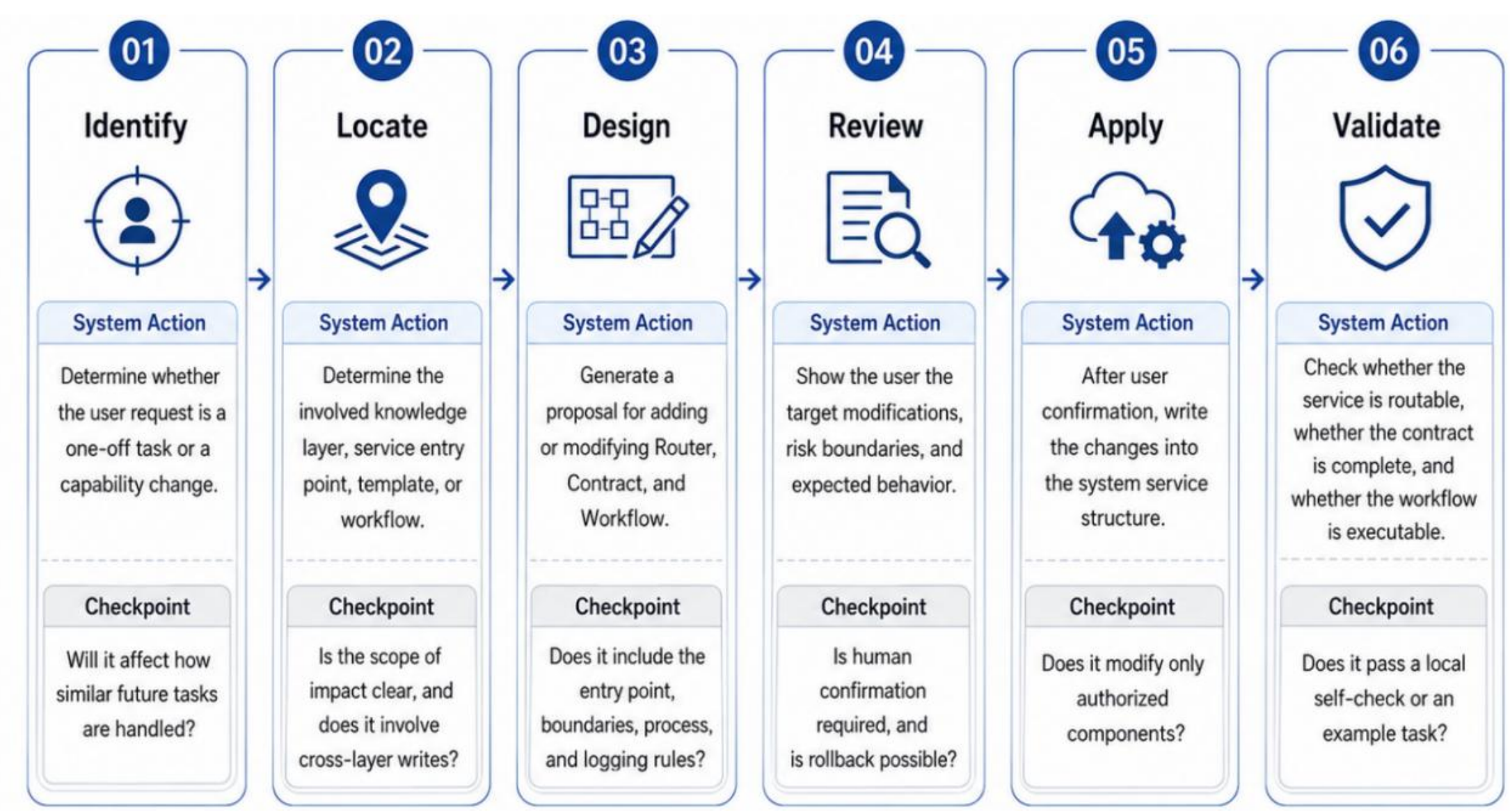

*Figure 9.1 The governed service-evolution process in Valhalla*

A service-evolution instance typically comprises requirement classification, candidate-component generation, boundary checking, human confirmation, registration, validation, and auditing. Unlike a one-off task, a service change affects subsequent requests of the same type; therefore, completion of the current task cannot serve as the sole criterion for successful registration. Research on skill lifecycles provides a reference for creating and revising externalized capabilities [10-14], while Valhalla further emphasizes the objects being changed and the scope of knowledge writes.

## 9.3 Governance Boundaries

The principal risks of service evolution include implicit inheritance, permission expansion, cross-layer writes, and insufficient validation. Valhalla adopts component independence, explicit scope, non-inheritance of confirmation, and diagnosable failure as design constraints. These constraints can become enforceable guarantees only when supported by external validators, tests, and orchestration mechanisms; when specified solely in a Contract or Workflow, they may still be misinterpreted by the model [22-23].

| **Design constraints for governed service evolution** |
| --- |
| The source mapping of an Entity must be retained through a registered Resource and must not be overwritten with content lacking provenance. |
| Each new writable Workflow must be bound to an explicit Contract, with its permissions and risks declared separately. |
| The output of a one-off task must not be registered directly as a long-term system service without review. |
| Cross-layer effects must be declared explicitly; operations requiring confirmation must not inherit authorization from other stages. |
| Service extensions should record changes and retain paths for termination, correction, and recovery in the event of validation failure. |

## 9.4 Example: Adding an "Experimental-Protocol Review" Service

The basic version of Valhalla does not include an experimental-protocol review system service; users must evolve such a service according to their own requirements.

Suppose a research group submits the following request: "We need an experimental-protocol review workflow that checks whether a protocol covers the key variables, whether its methodological choices are consistent with existing evidence, and whether it contains unverified assumptions." One practical approach is to use an LLM to formulate a corresponding review procedure, or to adopt an established procedure and have Valhalla evolve in accordance with it. In this way, the resulting Valhalla system service can use both the established procedure and the structured knowledge governed by Valhalla. Valhalla first classifies the request as a capability change rather than one-off text generation.

The system then determines that this capability primarily reads Entity, Relationship, and Graph objects and may cite Resource objects as evidence sources when necessary, but should not directly modify original materials in the File layer. It accordingly adds a Router entry so that similar requests can enter the "Experimental-Protocol Review" service. It then defines a Contract specifying that the service may read the Entity, Relationship, and Graph objects relevant to the current project and may output a review report and a list of hypotheses requiring validation, but may not modify the states of existing relationships without confirmation. Finally, it creates a Workflow that specifies the review steps, including identifying experimental objectives, mapping relevant entities, examining relationship chains, locating evidence gaps, generating risk alerts, and recording the review log.

This example demonstrates that service evolution operates on identifiable components rather than on the system as a whole. Whether the new capability can be executed reliably still depends on testing, input materials, and domain review; generating the service files alone does not constitute evidence of the capability's effectiveness.

### 9.5 Complementarity of Automatic Generation and Manual Maintenance

Researchers may always add or modify Router, Contract, and Workflow components manually. The value of natural language and lightweight structured representations lies in enabling domain specialists to participate in defining knowledge objects and reviewing processes, rather than placing system maintenance entirely in the hands of the model.

Model generation can reduce initial setup costs, whereas manual maintenance supports fine-grained adjustment to specific requirements.

In one case involving the extension of a literature-ingestion system service for antibody design, a candidate service triggered a safety review because of potentially high-risk biological information. Subsequent manual revision restricted the scope to review-level background information, conceptual conclusions, evaluation metrics, and non-operational risk descriptions, while excluding executable wet-lab procedures and high-risk optimization pathways.

This process illustrates that human intervention can narrow service boundaries, but it neither demonstrates that the safety classifications of the underlying model are stable nor constitutes a general compliance guarantee for all biomedical tasks.

## 10 Case Study: A Knowledge-Governed Scientific Writing Workflow Based on Valhalla

### 10.1 Experimental Objectives

As discussed above, existing LLM agents have demonstrated that structured knowledge representations, tool use, and task planning can improve the automation of complex scientific tasks [15,20,25-27]. In real research teams, however, long-term scientific work requires more than the completion of individual tasks: it also requires the continual accumulation of domain knowledge, maintenance of evidential relationships, and ongoing adaptation of analytical workflows as research directions evolve.

Valhalla is not intended to re-establish the value of structured knowledge for scientific agents. Rather, it explores, on that basis, a long-term intelligent work environment for research teams. This case study therefore examines three core design hypotheses underlying Valhalla:

(1) Structured knowledge states can transform scientific knowledge from one-off document inputs into knowledge assets that can be continuously maintained and reused.

(2) A Router–Contract–Workflow service architecture can enable research teams to extend domain-specific task capabilities with a low barrier to entry.

(3) A governed evolution mechanism can preserve the transparency and maintainability of an LLM-agent system throughout long-term use.

To examine these hypotheses, we selected a cross-domain scientific writing task as the case study and evaluated Valhalla's integrated support for knowledge ingestion, system evolution, and the generation and review of scientific text.

## 10.2 Experimental Task and Workflow Design

We used PVRIG/antibody review generation as the experimental case [28-29].

The task spans:

- tumor immunology mechanisms;
- immune-checkpoint networks;
- antibody engineering design;
- optimization of Fc functions; and
- translational evaluation.

The topic encompasses molecular mechanisms, biomedical knowledge, and engineering design strategies and is therefore representative of cross-domain scientific writing [28-29].

The experiment comprised three stages:

1. evaluation of domain-knowledge ingestion and system-evolution capabilities;
2. evaluation of knowledge-asset sharing and structured organization; and
3. evaluation of knowledge-state-based support for scientific writing.

### 10.2.1 Evaluation of Valhalla's System-Evolution Capability

Because research domains differ in their knowledge-organization practices and evaluation criteria, general-purpose scientific agents often cannot directly accommodate the requirements of a new domain. We therefore first examined whether Valhalla could extend its system services in response to domain-specific requirements.

We began by developing a domain-knowledge ingestion specification tailored to papers on antibody design. The specification defined:

- how knowledge entities should be organized;
- which experimental information should be retained;
- requirements for recording evidence; and
- the boundaries of domain-relevant knowledge.

The domain specification was then provided to Valhalla, which used its bootstrapped evolution mechanism to generate a new domain-knowledge ingestion service.

The process comprised:

1. generating a new service design from the domain requirements;
2. creating a corresponding Router entry so that the system could recognize ingestion tasks in this domain;
3. defining a Contract that constrained the service's inputs, outputs, and scope of knowledge modification; and
4. writing a Workflow that specified paper parsing, Entity extraction, evidence mapping, and logging.

This process produced a dedicated knowledge-ingestion service for the antibody-design domain: antibody_design_ingest.

The service was subsequently distributed to the researchers participating in the experiment, enabling different members to ingest papers from the domain in parallel.

During use, some materials triggered the LLM's safety restrictions. In response, the researchers manually evolved the service by modifying its Workflow and adding domain-specific safety constraints. The revised service was limited to conceptual knowledge, evaluation metrics, research background, and publicly reported experimental results, while excluding high-risk operational content unsuitable for automated generation.

Ingestion of the remaining materials continued after these constraints had been adjusted.

#### 10.2.2 Evaluation of Valhalla's Knowledge-Asset Sharing Capability

To examine whether Valhalla could support team-level knowledge sharing, we integrated knowledge objects independently ingested by different team members.

The resulting knowledge base contained:

- 26 paper Resources;
- 80 Entities; and
- 92 Relationships.

On average, each paper yielded approximately three knowledge entities.

Valhalla's relationship-construction service was then used to organize the semantic relationships among Entities, including:

- mechanistic relationships;
- methodological dependencies;
- evidential support relationships; and
- method-evaluation relationships.

This process produced a knowledge graph for the research topic.

The purpose of this stage was to examine whether knowledge produced by different team members could be separated from their individual contexts and transformed into a shared, transferable structured knowledge state.

### 10.2.3 Evaluation of Valhalla's Support for Scientific Writing

To assess the effect of structured knowledge states on scientific writing, we designed two comparative workflows.

**Baseline workflow**

1. Clear the LLM agent's conversation history.
2. Conduct the review-writing task:
   (1) load the academic-paper-orchestrator skill;
   (2) use the 26 ingested papers as references; and
   (3) use an existing review as a structural template to generate *PVRIG Antibody Design Review 1*.
3. Review the paper using Valhalla's paper_review service.
4. Revise the paper using the paper_polish service on the basis of the review results.

The final output was *PVRIG Antibody Design Review 1_Revised and Polished*.

**Valhalla workflow**

1. Clear the LLM agent's conversation history.
2. Conduct the review-writing task:
   (1) start Valhalla's paper_orchestrator service, a system service that reproduces the academic-paper-orchestrator skill while providing access to the knowledge base;
   (2) access the completed knowledge base; and
   (3) use the same review template to generate *PVRIG Antibody Design Review 2*.
3. Review the paper using the paper_review service.
4. Revise the paper using the paper_polish service in conjunction with the knowledge base.

The final output was *PVRIG Antibody Design Review 2_Revised and Polished*.

## 10.3 Experimental Results

### 10.3.1 Execution Time

The time required by the two workflows is shown below:

| Experimental step | Duration (min) |
|---|---|
| Writing 1 | 47.5 |
| Review 1 | 3.5 |
| Revision 1 | 12.5 |
| Writing 2 | 10 |
| Review 2 | 11.5 |
| Revision 2 | 10.5 |

The initial generation time was substantially lower in the second scientific writing workflow, for which an existing knowledge state and corresponding system services were available.

### 10.3.2 Evaluation of Paper Quality

1. The generated outputs were evaluated by domain experts.

The evaluation indicated that the review generated with the Valhalla knowledge base provided more comprehensive knowledge coverage, more specific references to experimental results, and more in-depth discussion of mechanisms.

Compared with the output generated solely through the paper-writing workflow, the Valhalla-supported text drew more extensively on experimental data and research conclusions from the source papers.

2. ChatGPT was additionally used for an auxiliary evaluation, with the following results:

| Experimental step | Duration (min) |
|---|---|
| Writing 1 | 47.5 |
| Review 1 | 3.5 |
| Revision 1 | 12.5 |
| Writing 2 | 10 |
| Review 2 | 11.5 |
| Revision 2 | 10.5 |

Detailed evaluation:

| Evaluation dimension | Review 1 | Revised Review 1 | Review 2 | Revised Review 2 |
|---|---|---|---|---|
| Depth of scientific content (30) | 26 | 27 | 28 | 29 |
| Logical structure of the review (25) | 20 | 23 | 23 | 24 |
| Rigor of evidence (20) | 15 | 19 | 18 | 19 |
| Expertise in antibody engineering (15) | 12 | 12 | 14 | 14 |
| Innovative framing and publication potential (10) | 9 | 9 | 8 | 8 |
| **Total score** | **82** | **90** | **91** | **94** |

## 10.4 Analysis of Results

The results suggest that Valhalla's principal advantage does not arise from an improvement in one-off text generation. Instead, it stems from the long-term organization of scientific knowledge states, operational workflows, and system capabilities.

First, with respect to knowledge state, Valhalla transforms source papers into structured knowledge objects containing entities, relationships, and evidence mappings, thereby reducing the scientific writing process's reliance on one-off context inputs.

In the conventional workflow, the LLM must repeatedly identify conceptual relationships and experimental evidence across a large collection of papers. In the Valhalla workflow, relevant knowledge has already been organized through Entities and Relationships, allowing the model to construct arguments around stable knowledge objects. The generated output was consequently stronger in its use of experimental evidence, explanation of mechanisms, and structural organization.

This finding suggests that the principal bottleneck for a scientific agent is not limited to language-generation capability; it also concerns whether the agent possesses a knowledge state that can be maintained over time.

Second, with respect to the extension of system capabilities, the experiment demonstrated that Valhalla's service-evolution mechanism can support domain adaptation.

Conventional agent systems generally rely on tools and workflows predefined by developers. Moving into a new research domain therefore requires the redevelopment of system functionality. Valhalla instead encapsulates system capabilities as Router, Contract, and Workflow components, allowing researchers to generate new service structures in response to domain requirements and to adjust their behavioral boundaries through human review.

The objects of system evolution are therefore not opaque software code, but readable, modifiable, and governable service components.

Third, with respect to scientific writing quality, the results indicate that support from the knowledge base primarily improved the depth of scientific content, logical structure, and rigor of evidence rather than merely improving linguistic expression.

This result suggests that structured knowledge states benefit scientific agents by reducing knowledge loss and breaks in evidential chains, thereby enabling the model to organize new scientific narratives on the basis of existing research relationships.

Finally, the experiment illustrates how Valhalla differs from conventional RAG systems and single-task agents.

Conventional methods primarily optimize retrieval and generation within an individual task, whereas Valhalla focuses on the joint accumulation of knowledge, task workflows, and system capabilities over the course of long-term scientific work. The knowledge base, domain-ingestion service, and paper-review workflow produced in this experiment can all be reused in subsequent tasks; the system's value may therefore increase with continued use.

## 10.5 Efficiency Analysis

In addition to improvements in paper quality, the experiment revealed changes in the efficiency of scientific-task execution.

Compared with the baseline generation workflow, the workflow supported by Valhalla's knowledge state and system services:

- reduced initial generation time by approximately 79%; and
- kept a single scientific writing iteration to approximately 30 minutes.

It should be noted that constructing the knowledge state incurred additional costs. Initial knowledge ingestion required approximately 300 minutes in this experiment; the complete workflow may therefore offer no time advantage if only one task is performed.

For long-term scientific work, however, knowledge ingestion constitutes a one-time investment in asset construction. When the same knowledge base is subsequently used for paper writing, experimental design, review updates, or project discussions, the initial cost can be amortized across multiple tasks.

Valhalla's efficiency advantage therefore does not arise from accelerating a single task, but from reusing knowledge and workflows throughout a long-term scientific process.

# 11 Limitations, Failure Modes, and Governance Boundaries

The current evidence for Valhalla is derived primarily from its architectural description and a single case study. Its limitations include not only insufficient internal and external validity, but also engineering risks associated with knowledge quality, natural-language constraints, service coupling, external safety policies, state consistency, and context management. These issues must be stated separately from the design intent to avoid presenting normative rules as runtime guarantees.

## 11.1 Scope of Evidence and Internal Validity

The case study in Section 10 included no randomization, independent replication, strong baseline under a matched resource budget, component ablation, or auditable expert evaluation. The model-assisted scoring also did not disclose the complete judge configuration or tests for bias [30]. The differences in time and scores therefore cannot support causal attribution, and D1–D3 have not yet been empirically validated. Future work will evaluate them through larger-scale and more rigorous experiments.

Constructing a knowledge state requires resource registration, provenance mapping, object extraction, relationship construction, and quality control. These initial costs may not be recovered in one-off or infrequent tasks. For long-term projects, the benefits of reuse must be measured through ingestion, maintenance, retrieval, and handover records collected at multiple time points rather than inferred solely from the system's design for reusability.

## 11.2 Risks Arising from the LLM's Underlying Rules

One issue identified to date concerns batch ingestion. Even when a Workflow requires documents to be processed individually, the model may combine multiple documents in a single analysis, thereby mixing methods, conditions, and conclusions across sources. The model may interpret such combination as an efficiency optimization, but it undermines evidence mapping in which each individual source serves as the boundary.

## 11.3 Natural-Language Governance and Service Coupling

Router, Contract, and Workflow are represented in natural language and lightweight structured formats. Although this makes them readable to domain researchers, it also permits the model to reinterpret or reconstruct the rules. The model may implement an instruction to “refer to a service” by directly inheriting or invoking that service, thereby creating implicit dependencies between services that are intended to remain independent.

Such coupling may have no immediate effect on execution, but it expands the scope of subsequent modifications: when the original service is updated, the behavior of the new service may change accordingly, even though the dependency has not been explicitly registered. Similar risks apply to the transfer of permissions and confirmations; the qualifications of one service must not be inherited by a subsequent service by default.

The following boundary must therefore be made explicit whenever a service is added or modified:

A new service may refer to the structure and design patterns of an existing service, but it must not inherit, invoke, or depend on the existing service’s specific implementation unless this is declared. All dependencies, read/write scopes, and confirmation conditions must be registered and verified separately.

## 11.4 External Safety Policies and Partial Commits

Valhalla runs on general-purpose language models and their platforms and is therefore constrained by both its own service contracts and the underlying model’s safety policies. Tasks involving drug discovery, biomedical analysis, or experimental design may trigger platform-level safety review.

Platform review is necessary to reduce high-risk uses, but its decisions occur outside Valhalla’s governance system. Even after a task has passed internal permission checks, the underlying model may still refuse to continue, resulting in failure.

Valhalla must not bypass the underlying safety mechanisms. A more practical approach is to use staged commits, failure markers, human review, and recoverable transactions, while clearly distinguishing among a “platform refusal,” a “service failure,” and a “committed knowledge state.” The current implementation does not yet provide these mechanisms in full.

## 11.5 State Consistency, Snapshots, and Rollback

Resources, knowledge objects, relationships, graphs, and services persist as directly editable file states. This lowers the barrier to human review but also creates tangible risks of partial commits, concurrent overwrites, and inadvertent modification. The current version does not yet bind every operation to an automatic snapshot, transactional commit, and verifiable rollback.

At present, recovery relies primarily on local backups and manual restoration. Future versions must establish change sets, pre-commit validation, atomic replacement, failure rollback, and recovery exercises for high-risk writes, while recording who modified which objects, when the modifications occurred, and on the basis of which confirmation. Without these mechanisms, “governable” can signify only that the process is visible; it cannot be equated with transactional safety.

### 11.6 Context Management, Evaluation Reliability, and External Validity

Loading Router, Contract, and Workflow in stages can reduce the amount of irrelevant material occupying the context, but the current version has no independent context-scheduling layer. The system cannot precisely determine when particular rules, resources, confirmations, and task states should be loaded, compressed, evicted, or restored [24].

Long-running tasks still depend on the underlying model's context window and automatic compression. Earlier constraints may be summarized, provenance identities may lose detail, and stage-specific confirmations may become detached from the current operation. Because internal platform compression policies are generally unobservable, the causes of failure can be difficult to reconstruct.

An independent context-management layer requires, at minimum, the following capabilities:

distinguishing system rules, task states, knowledge objects, and temporary reasoning results;

assigning priorities, lifecycles, and loading conditions to different context objects;

loading the Contract, resources, and execution records required for each Workflow stage;

persisting structured task states before compression rather than relying entirely on natural-language summaries;

marking critical constraints, user confirmations, and provenance identities as content that must be retained or reloaded;

evicting irrelevant content and loading the state required for the next stage when the task transitions between stages; and recording the rules and resources actually loaded at each step to support execution tracing and error analysis.

In addition, the model-assisted evaluation in the current case study included no independent blinded assessment, repeated judges, bias testing, or measures of inter-rater agreement and therefore cannot serve as a stable measure of quality [30]. External validity is also limited by the use of a single topic, a single team, and a single run. Future research should jointly evaluate knowledge quality, task performance, governance failures, long-term cost, and cross-member transfer, and should disclose configurations and raw records in a form that permits independent review.

## 12 Conclusion

This paper has presented Valhalla, a layered knowledge-state and service-governance framework for long-term scientific knowledge work. The data plane uses File, Resource, Entity, Relationship, and Graph to distinguish document entry points, stable identities, knowledge objects, semantic judgments, and task-oriented views, respectively. The control plane uses Router, Contract, and Workflow to organize request classification, eligibility checks, and constrained execution.

The framework's primary contribution is a layered architectural protocol that addresses a missing dimension of current structured-knowledge systems. This protocol makes structured knowledge a shareable and recomposable asset, extending LLM-assisted scientific collaboration from the sharing of agents to the sharing of structured knowledge.

The PVRIG writing case in Section 10 shows that the prototype can construct layered knowledge states and invoke relevant services, but its time and quality scores are derived from a single, independently unaudited workflow. The available evidence does not demonstrate that Valhalla outperforms RAG, GraphRAG, general memory systems, or workflow-only methods, nor does it establish long-term reuse benefits or the realization of D1–D3 [4-6,18-19].

Future work should conduct multi-task comparisons, component ablations, independent replications, blinded expert evaluations, knowledge-quality audits, fault injection, and longitudinal cost measurements under a frozen protocol. It should also provide independently reviewable code, configurations, corpus inventories, and raw evaluation records. Only after these evaluations have been completed will it be possible to determine whether layered knowledge states and service governance constitute a generalizable infrastructure for scientific agents.

# References

[1] HOGAN A, BLOMQVIST E, COCHEZ M, et al. Knowledge Graphs[J]. ACM Computing Surveys, 2021, 54(4): 1-37. DOI: 10.1145/3447772.

[2] WILKINSON M D, DUMONTIER M, AALBERSBERG I J, et al. The FAIR Guiding Principles for Scientific Data Management and Stewardship[J]. Scientific Data, 2016, 3: 160018. DOI: 10.1038/sdata.2016.18.

[3] MOREAU L, MISSIER P, eds. PROV-DM: The PROV Data Model[EB/OL]. W3C Recommendation, (2013-04-30)[2026-07-25]. https://www.w3.org/TR/2013/REC-prov-dm-20130430/.

[4] LEWIS P, PEREZ E, PIKTUS A, et al. Retrieval-Augmented Generation for Knowledge-Intensive NLP Tasks[C/OL]//Advances in Neural Information Processing Systems 33. Red Hook: Curran Associates, 2020: 9459-9474[2026-07-25]. https://proceedings.neurips.cc/paper/2020/hash/6b493230-Abstract.html.

[5] EDGE D, TRINH H, CHENG N, et al. From Local to Global: A Graph RAG Approach to Query-Focused Summarization[EB/OL]. (2025-02-19)[2026-07-25]. https://arxiv.org/abs/2404.16130. DOI: 10.48550/arXiv.2404.16130.

[6] PACKER C, WOODERS S, LIN K, et al. MemGPT: Towards LLMs as Operating Systems[EB/OL]. (2024-02-12)[2026-07-25]. https://arxiv.org/abs/2310.08560. DOI: 10.48550/arXiv.2310.08560.

[7] PARK J S, O'BRIEN J C, CAI C J, et al. Generative Agents: Interactive Simulacra of Human Behavior[C]//Proceedings of the 36th Annual ACM Symposium on User Interface Software and Technology. New York: ACM, 2023: Article 2, 1-22. DOI: 10.1145/3586183.3606763.

[8] GUTIÉRREZ B J, SHU Y, GU Y, et al. HippoRAG: Neurobiologically Inspired Long-Term Memory for Large Language Models[C]//Advances in Neural Information Processing Systems 37. 2024: 59532-59569. DOI: 10.52202/079017-1902.

[9] GUTIÉRREZ B J, SHU Y, QI W, et al. From RAG to Memory: Non-Parametric Continual Learning for Large Language Models[C/OL]//Proceedings of the 42nd International Conference on Machine Learning. PMLR, 2025, 267: 21497-21515[2026-07-25]. https://proceedings.mlr.press/v267/gutierrez25a.html.

[10] WANG G, XIE Y, JIANG Y, et al. Voyager: An Open-Ended Embodied Agent with Large Language Models[J/OL]. Transactions on Machine Learning Research, 2024[2026-07-25]. https://openreview.net/forum?id=ehfRiF0R3a.

[11] ZHOU H, GUO S, LIU A, et al. Memento-Skills: Let Agents Design Agents[EB/OL]. (2026-03-19)[2026-07-25]. https://arxiv.org/abs/2603.18743. DOI: 10.48550/arXiv.2603.18743.

[12] YANG Y, LI J, PAN Q, et al. AutoSkill: Experience-Driven Lifelong Learning via Skill Self-Evolution[EB/OL]. (2026-03-05)[2026-07-25]. https://arxiv.org/abs/2603.01145. DOI: 10.48550/arXiv.2603.01145.

[13] LIN H, LI P, SONG J, JIANG F, ZHANG T. MUSE-Autoskill: Self-Evolving Agents via Skill Creation, Memory, Management, and Evaluation[EB/OL]. (2026-07-03)[2026-07-25]. https://arxiv.org/abs/2605.27366. DOI: 10.48550/arXiv.2605.27366.

[14] ZHOU C, CHAI H, CHEN W, et al. Externalization in LLM Agents: A Unified Review of Memory, Skills, Protocols and Harness Engineering[EB/OL]. (2026-04-09)[2026-07-25]. https://arxiv.org/abs/2604.08224. DOI: 10.48550/arXiv.2604.08224.

[15] GOTTWEIS J, WENG W H, DARYIN A, et al. Accelerating Scientific Discovery with Co-Scientist[J]. Nature, 2026, 655: 487-496. DOI: 10.1038/s41586-026-10644-y.

[16] MI Y, GAO Z, MA X, LI Q. Building LLM Agents by Incorporating Insights from Computer Systems[EB/OL]. (2025-04-06)[2026-07-25]. https://arxiv.org/abs/2504.04485. DOI: 10.48550/arXiv.2504.04485.

[17] ZHU D, ZHENG L N, CHEN Z. FundaPod: A Multi-Persona Agent Pod Platform with Knowledge Graph Memory for AI-Assisted Fundamental Investment Research[EB/OL]. (2026-06-18)[2026-07-25]. https://arxiv.org/abs/2605.27864. DOI: 10.48550/arXiv.2605.27864.

[18] SOUZA R, POTEET T, ETZ B, et al. LLM Agents for Interactive Workflow Provenance: Reference Architecture and Evaluation Methodology[C]//Proceedings of the SC ’25 Workshops of the International Conference for High Performance Computing, Networking, Storage and Analysis. New York: ACM, 2025: 2257-2268. DOI: 10.1145/3731599.3767582.

[19] JONNALAGEDDA P, YAO Y, GAO X, et al. Executable Schema Contracts: From Automatic Ingestion to Multi-Source Retrieval[EB/OL]. (2026-06-03)[2026-07-25]. https://arxiv.org/abs/2606.05415. DOI: 10.48550/arXiv.2606.05415.

[20] WANG Z, CHEN Z, YANG Z, et al. Empowering Biomedical Evidence Exploration and Synthesis with Deep Knowledge Graph Research[J]. Nature Machine Intelligence, 2026, 8: 1142-1156. DOI: 10.1038/s42256-026-01266-0.

[21] OBSIDIAN. Graph View[EB/OL]. [2026-07-25]. https://help.obsidian.md/plugins/graph.

[22] TABASSI E. Artificial Intelligence Risk Management Framework (AI RMF 1.0)[R]. Gaithersburg: National Institute of Standards and Technology, 2023. DOI: 10.6028/NIST.AI.100-1.

[23] SALTZER J H, SCHROEDER M D. The Protection of Information in Computer Systems[J]. Proceedings of the IEEE, 1975, 63(9): 1278-1308. DOI: 10.1109/PROC.1975.9939.

[24] LIU N F, LIN K, HEWITT J, et al. Lost in the Middle: How Language Models Use Long Contexts[J]. Transactions of the Association for Computational Linguistics, 2024, 12: 157-173. DOI: 10.1162/tacl_a_00638.

[25] BRAN A M, COX S, SCHILTER O, et al. Augmenting Large Language Models with Chemistry Tools[J]. Nature Machine Intelligence, 2024, 6: 525-535. DOI: 10.1038/s42256-024-00832-8.

[26] BOIKO D A, MACKNIGHT R, KLINE B, et al. Autonomous Chemical Research with Large Language Models[J]. Nature, 2023, 624: 570-578. DOI: 10.1038/s41586-023-06792-0.

[27] LU C, LU C, LANGE R T, et al. The AI Scientist: Towards Fully Automated Open-Ended Scientific Discovery[EB/OL]. (2024-09-01)[2026-07-25]. https://arxiv.org/abs/2408.06292. DOI: 10.48550/arXiv.2408.06292.

[28] WHELAN S, OPHIR E, KOTTURI M F, et al. PVRIG and PVRL2 Are Induced in Cancer and Inhibit CD8+ T-Cell Function[J]. Cancer Immunology Research, 2019, 7(2): 257-268. DOI: 10.1158/2326-6066.CIR-18-0442.

[29] XUE H, ZHANG Z, LI L, et al. Characterization of a Novel Anti-PVRIG Antibody with Fc-Competent Function That Exerts Strong Antitumor Effects via NK Activation in Preclinical Models[J]. Cancer Immunology, Immunotherapy, 2024, 73(5): 81. DOI: 10.1007/s00262-024-03671-z.

[30] WANG P, LI L, CHEN L, et al. Large Language Models Are Not Fair Evaluators[C]//Proceedings of the 62nd Annual Meeting of the Association for Computational Linguistics: Volume 1, Long Papers. Bangkok: Association for Computational Linguistics, 2024: 9440-9450. DOI: 10.18653/v1/2024.acl-long.511.